\documentclass[reprint,prl,amsmath,amssymb,superscriptaddress,showpacs]{revtex4-1}
\usepackage{hyperref}
\usepackage{graphics}
\usepackage{enumerate}
\usepackage{hyperref}

\usepackage{color}

\newcommand{\ket}[1]{\left|{#1}\right\rangle}
\newcommand{\braket}[1]{\left\langle{#1}\right\rangle}

\usepackage{ulem}

\begin{document}

\title{Charge-insensitive single-atom spin-orbit qubit in silicon}
\author{Joe Salfi}
\affiliation{School of Physics, The University of New South Wales, Sydney, NSW 2052, Australia.}
\affiliation{Centre for Quantum Computation and Communication Technology, The University of New South Wales, Sydney, NSW 2052, Australia.}
\author{Jan A. Mol}
\affiliation{School of Physics, The University of New South Wales, Sydney, NSW 2052, Australia.}
\affiliation{Centre for Quantum Computation and Communication Technology, The University of New South Wales, Sydney, NSW 2052, Australia.}
\author{Dimitrie Culcer}
\affiliation{School of Physics, The University of New South Wales, Sydney, NSW 2052, Australia.}
\author{Sven Rogge}
\affiliation{School of Physics, The University of New South Wales, Sydney, NSW 2052, Australia.}
\affiliation{Centre for Quantum Computation and Communication Technology, The University of New South Wales, Sydney, NSW 2052, Australia.}
\date{\today}


\begin{abstract}
High fidelity entanglement of an on-chip array of spin qubits poses many challenges. Spin-orbit coupling (SOC) can ease some of these challenges by enabling long-ranged entanglement via electric dipole-dipole interactions, microwave photons, or phonons.  However, SOC exposes conventional spin qubits to decoherence from electrical noise.  Here we propose an acceptor-based spin-orbit qubit in silicon offering long-range entanglement at a sweet spot where the qubit is protected from electrical noise. The qubit relies on quadrupolar SOC with the interface and gate potentials.  As required for surface codes, $10^5$ electrically mediated single-qubit and $10^4$ dipole-dipole mediated two-qubit gates are possible in the predicted spin lifetime.  Moreover, circuit quantum electrodynamics with single spins is feasible, including dispersive readout, cavity-mediated entanglement, and spin-photon entanglement. An industrially relevant silicon-based platform is employed.
\end{abstract}

\maketitle

In recent years, the coherence and control fidelity of solid-state qubits has dramatically improved\cite{Majer:2007em,Shulman:2012fk,Barends:2014fu,Waldherr:2014kt,Veldhorst:2015jea} and spin qubits\cite{Loss:1998ia,Kane:1998ce,Petta:2005kn} with highly desirable properties have been demonstrated.\cite{Muhonen:2014bj,Veldhorst:2014eq} However, many obstacles remain to efficiently entangle a large array of spin qubits on a chip.  For example, exchange is inherently vulnerable to decoherence from electrical fluctuations\cite{Hu:2006ih,Culcer:2009fq,Dial:2013cb}, coupling spin to charge noise.  Minimizing decoherence and improving control in the face of noise is the key issue for large-scale quantum computing, because it ultimately determines if the error-correction resources can be managed for a large qubit array.\cite{Fowler:2012fi}  Moreover, exchange-based entanglement is inherently short-ranged, making fabrication challenging for gates in quantum dot arrays\cite{Loss:1998ia}, and placing strict demands on Si:P donor placement.\cite{Kane:1998ce} 

Here we propose a single-acceptor spin-orbit qubit where the unique properties of hole spins give a host of desirable attributes.  First, spin-orbit coupling (SOC) enables long-ranged entanglement via microwave photons or electric dipole-dipole interactions\cite{Golovach:2006hx,Flindt:2006dn,Trif:2007fx,Trif:2008cb,Bulaev:2005jz,Bulaev:2007hj,Palyi:2012hj,Kloeffel:2013iv,Ruskov:2013kq,Hu:2012bt,Taylor:2013cq}, of interest for hybrid quantum systems\cite{Blais:2004kn,Wallraff:2004dy,Petersson:2012cv,Xiang:2013hm,Viennot:2015ir}, improving error correction\cite{JochymOConnor:2015wp}, and reducing fabrication demands compared with exchange coupled schemes. Second, and most remarkably, we find a sweet spot where coherence is insensitive to electrical noise and electric dipole spin resonance\cite{Nowack:2007du,NadjPerge:2010kw,Medford:2013eka} (EDSR) is maximized.  Consequently, coherence and gate timings are protected from electrical noise at the Hamiltonian level, and one- and two-qubit gate times are optimized. In comparison, electric field noise dephases conventional spin-orbit qubits\cite{Huang:2014vf,Bermeister:2014jb} and acceptor charge qubits.\cite{Golding:2003tq,Ruskov:2013kq}  The coherence of our spin-orbit qubit benefits from reduced hyperfine coupling of holes\cite{Chekhovich:2012ks} and $^{28}$Si enrichment\cite{Tyryshkin:2011fi}, and has much longer phonon relaxation times than acceptor charge qubits.\cite{Golding:2003tq,Ruskov:2013kq}  Finally, the acceptors naturally confine single holes that can be manipulated in silicon nanoelectronic devices\cite{vanderHeijden:2014fp}.  

The exceptional properties of the qubit derive from the quadrupolar SOC\cite{Winkler:2004ka,Culcer:2006jc,Culcer:2007gs,Winkler:2008bc} contained in the spin-3/2 Luttinger Hamiltonian\cite{Luttinger:1955ee} and in the interaction with the inversion asymmetric interface potential, not studied previously for acceptors.  This SOC is unusually strong for acceptors because it acts directly on the low-energy spin manifold, contrasting its indirect role in hole quantum dots.\cite{Bulaev:2005jz,Bulaev:2007hj,Zwanenburg:2009jr,Hu:2011ic,Pribiag:2013if,Voisin:2016jc} The SOC must be considered non-perturbatively to obtain the sweet spot, and the interface strongly enhances EDSR relative to a bulk acceptor. We find 0.2 ns one-qubit gate times, charge-noise immunity, and long phonon relaxation times at the sweet spot, allowing for $>10^5$ operations in the coherence time. Two-qubit entanglement based on spin-dependent electric dipole-dipole interactions\cite{Golovach:2006hx,Flindt:2006dn,Trif:2007fx} is feasible with $\sqrt{\textrm{SWAP}}$ times of $2$ ns, and $10^4$ operations in the coherence time.  EDSR also enables circuit quantum electrodynamics\cite{Blais:2004kn,Wallraff:2004dy,Petersson:2012cv,Xiang:2013hm,Viennot:2015ir} (cQED) with single-spin dispersive readout, and long distance spin-spin entanglement with $\sqrt{\textrm{SWAP}}$ times of $200$ ns.  Resonant spin-photon coupling with $g_c=5$ MHz is also feasible.   

\begin{figure}
\includegraphics{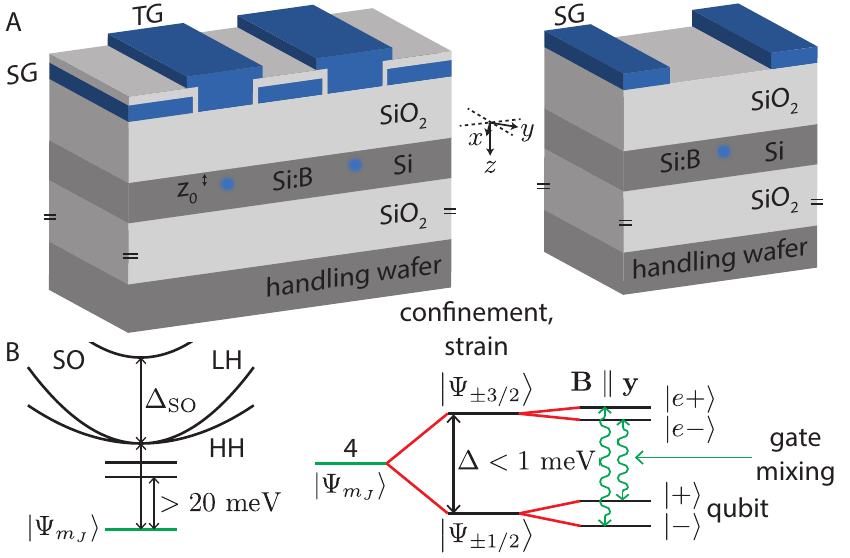}
\caption{A. Device schematic, showing near-interface Si:B impurity with gates to SG and TG to apply in-plane and vertical electric fields, respectively (left), or for cQED, gates forming the resonator apply both the in-plane and vertical electric fields (right).  An in-plane applied magnetic field ensures a long photon lifetime in the superconductor resonator. B. Electronic structure of an acceptor, where the splitting $\Delta$ is determined by the strain, interface, and gate field $E_z$.  Shown: $pE_z$-induced mixing of states in the $4\times4$ manifold due to the $T_d$ symmetry in the unit cell of the ion.  Not shown: LH-HH coupling from in-plane drive fields. }
\label{fdevice}
\end{figure}

{\it Qubit Concept.} The qubit is a hole spin bound to a single Si:B dopant\cite{vanderHeijden:2014fp,Mol:2015im,Salfi:2016uma}, implanted\cite{Morello:2010ga} or placed by scanning tunneling microscopy\cite{Fuechsle:2012bl,Miwa:2013ib} near an interface, in a strained silicon-on-insulator (SOI) substrate (Fig. \ref{fdevice}A). The key quadrupolar interactions, associated with interface inversion asymmetry and products $\{J_i,J_j\}=(J_iJ_j+J_jJ_i)$ of spin-$3/2$ matrices where $i (j)=x,y,z$, originate from strong SOC in the valence band, and have no analog in the conduction band.\cite{Winkler:2004ka,Culcer:2006jc,Culcer:2007gs,Winkler:2008bc} This SOC acts on the $4\times4$ ground state manifold $\left|\Psi_{m_J}\right\rangle$, \textit{i.e.}, the $m_J=\pm \tfrac{3}{2}$ and $m_J=\pm\tfrac{1}{2}$ Kramers doublets composed mostly of $\left|J=\tfrac{3}{2},m_J\right\rangle$ Bloch states.\cite{Bir:1963cu} For Si:B they are well isolated by $\sim 20$ meV from orbital excited states and 46 meV from the valence band edge\cite{Pavlov:2014ju} (Fig.~\ref{fdevice}B). 

The key quadrupolar interactions include the acceptor hole spin-mixing that is linear in electric fields, $H_{\rm E, ion}=2p/\sqrt{3}(E_z\{J_x,J_y\}+\textrm{c.p})$, associated with $T_d$ symmetry in the central cell \cite{Bir:1963iy}. Here, $p=0.26$ D is known for Si:B\cite{Kopf:1992gj} (1 D = 0.021 e$\cdot$nm). An electric field $E_z$ further breaks the envelope function parity by mixing excited states outside the $\left|\Psi_{m_J}\right\rangle$ manifold.\cite{Bir:1963cu} Projected into the $\left|\Psi_{m_J}\right\rangle$ subspace, this interaction is governed by $H_E=b(J_z^2-\tfrac{5}{4}I)E_z^2 + (2d/\sqrt{3})(\{J_y,J_z\}E_yE_z +\{J_z,J_x\}E_zE_x)$, where $b$ and $d$ split and mix the doublets, respectively. We verified that this holds for triangular interface wells, using (i) a Schrieffer-Wolff transformation\cite{Schrieffer:1966hu,Winkler:2003dc} with higher excited states in the spherical spin-3/2 basis\cite{Baldereschi:1973im}, and (ii) numerical, non-perturbative Luttinger-Kohn (LK) calculations with explicit ion and interface well potentials\cite{Lawaetz:1971cm,Bernholc:1977bf}. We find a splitting $\Delta_W(E_z)=\Delta_{\rm if}+\Delta_G(E_z)$ (Fig.~\ref{fdevice}B), where $\Delta_{\rm if}$ from the interface is larger for shallower acceptors (in agreement with experiments\cite{Mol:2015im}), and $\Delta_G(E_z)\propto E_z$ increases with increasing field. Moreover, quadrupolar SOC combining inversion asymmetry and in-plane electric fields is governed by terms $\alpha(E_z)E_{x,y}\propto E_zE_{x,y}$ that replace $dE_zE_{x,y}$ in $H_{E}$. 

{\it Operating point and sweet spot.} Here we show that the qubit splitting $\hbar\omega$ (between $\ket{+}$ and $\ket{-}$, Fig.~1B) in an in-plane magnetic field $\hat{\mathbf{y}}B$ depends on the electric field $E_z$ applied by the gates (Fig.~1A), and at the sweet spot, $\hbar\omega$ is insensitive to electric-field noise $\delta\mathbf{E}$ in all directions. Including magnetic fields, strain $\Delta_\epsilon$ (Supplemental Material\cite{SMref}) and the interface well, but not in-plane electric fields, we find an operating point Hamiltonian, 
\begin{equation}
H_{\rm op}=\left(
\begin{array}{cccc}
\Delta(E_z) & -i\varepsilon_Z & i\tfrac{\sqrt{3}}{2}\varepsilon_Z & -ipE_z \\
i\varepsilon_Z & \Delta(E_z) & ipE_z & -i\tfrac{\sqrt{3}}{2}\varepsilon_Z \\
-i\tfrac{\sqrt{3}}{2}\varepsilon_Z & -ipE_z & 0 & 0\\
ipE_z & i\tfrac{\sqrt{3}}{2}\varepsilon_Z & 0 & 0
\end{array}
\right)
\end{equation}
in the basis $\{\ket{\Psi_{{-1/2}}},\ket{\Psi_{1/2}},\ket{\Psi_{-3/2}},\ket{\Psi_{3/2}}\}$, where $\varepsilon_Z=g_1\mu_BB$, $\mu_B$ is the Bohr magneton, $g_1=1.07$ is the Land\'{e} g-factor for Si:B.\cite{Kopf:1992gj}, and $\Delta(E_z)=\Delta_W(E_z)-\Delta_\epsilon$ is the splitting between the light and heavy holes.  The cubic g-factor\cite{Kopf:1992gj} $g_2\ll g_1$ is temporarily neglected.  

\begin{figure*}
\includegraphics{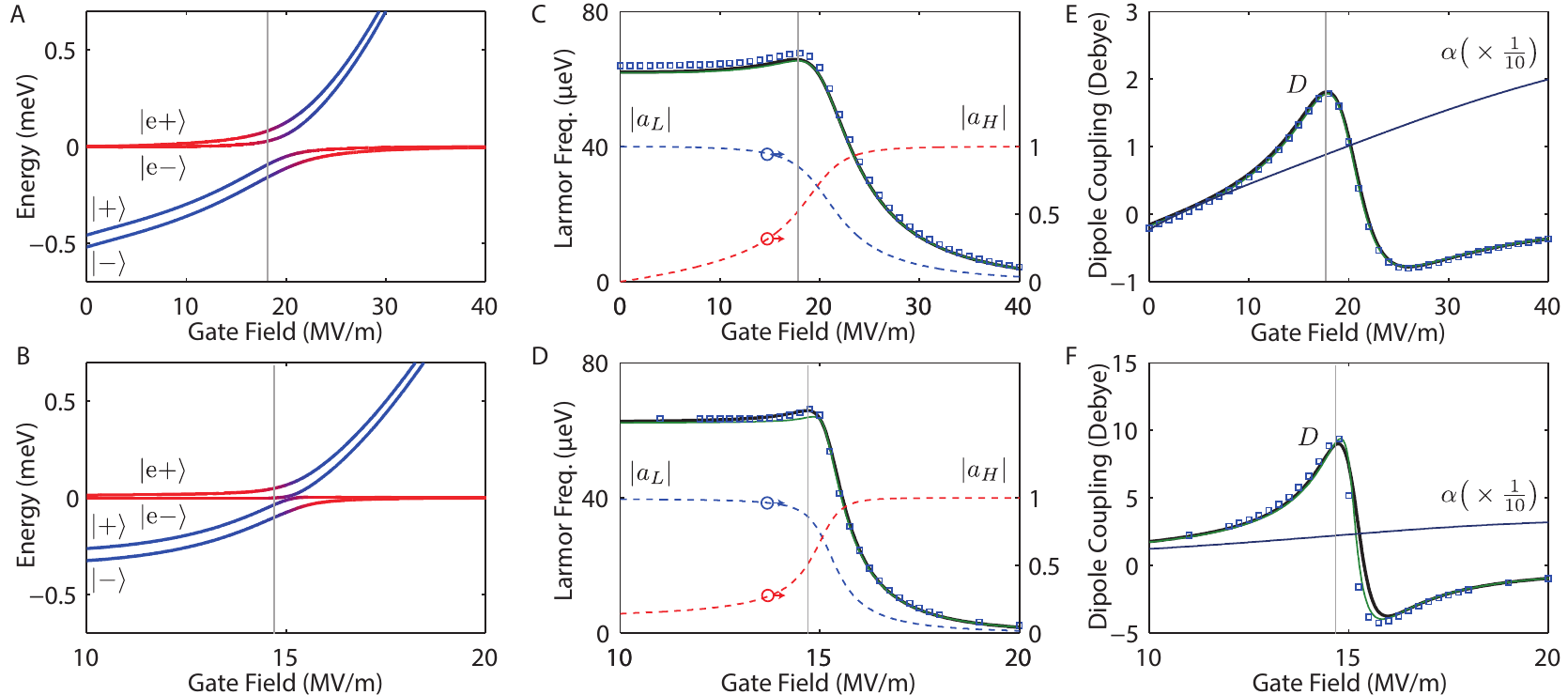}
\caption{Spin qubit levels $\varepsilon_{\pm}$ and $\varepsilon_{u\pm}$ for (A) $z_0=4.6$ nm and (B) $z_0=6.9$ nm, to zeroth order in $\lambda_{Zo}/(\varepsilon_u-\varepsilon_l)$. Qubit frequency for (C) $z_0=4.6$ nm and (D) $z_0=6.9$ nm using approximate (black), analytic (green), and full numerical (blue squares) models, in $B_0=0.5$ T. Spectral weights $|a_L|$ (blue dashed) and $|a_H|$ (red dashed) are shown.  EDSR coupling $D$ for (E) $z_0=4.6$ nm and (F) $z_0=6.9$ nm. We take $\Delta_{\epsilon}=0.62$ meV ($0.34$ meV) for $z_0=4.6$ nm ($6.9$ nm) achievable in SOI\cite{Celler:2003ha}, and exceeding disorder strain\cite{vanderHeijden:2014fp,Lo:2015kh}.  Parameters $\Delta_{\rm if}$, $\Delta_G(E_z)$, and $\alpha(E_z)$ were obtained non-perturbatively in a $6\times6$ LK basis including the cubic LK terms and the split-off holes.}
\label{fop}
\end{figure*}

Inspecting $H_{\rm op}$, $E_z$ mixes $\left|\Psi_{\pm1/2}\right\rangle$ and $\left|\Psi_{\mp 3/2}\right\rangle$ and these states have an avoided crossing when the interface well splitting compensates strain, \textit{i.e.}, $\Delta(E^0_z)=0$. In Fig.~1A we show that for appropriate strains $\Delta_\epsilon > \Delta_{\rm if}$, the anti-crossing can be obtained at $E^0_z\sim15$ MV/m for $z_0\sim 5$ nm acceptor depths.  

The field $E_z^0$ at such an anti-crossing is large enough that the level-repulsion gap $\Delta_{\rm gap}=2pE^0_z$ exceeds the Zeeman interactions, \textit{i.e.}, $\varepsilon_Z/\Delta_{\rm gap}\sim 0.1$.  This unusual aspect of our hole spin-orbit qubit \textit{c.f.} other proposals\cite{Golovach:2006hx,Flindt:2006dn,Trif:2007fx,Trif:2008cb,Bulaev:2005jz,Bulaev:2007hj} follows from the tunability of the spin-3/2 levels with strain and confinement, giving rise to the anti-crossing, and the strength of quadrupolar SOC\cite{Kopf:1992gj} relative to typical spin qubit Larmor frequencies.  We treat the quadrupolar SOC term $pE_z$ by a rotation that maps $pE_z$ exactly to the diagonal, to a basis $\{\ket{l-},\ket{l+},\ket{u-},\ket{u+}\}$ leaving Zeeman terms $\varepsilon_Z$ off-diagonal.  We obtain $\left|l\pm\right\rangle=a_L\left|\Psi_{\pm1/2}\right\rangle \pm i a_H\left|\Psi_{\mp 3/2}\right\rangle$, a low-energy Kramers pair (energy $\varepsilon_l=\tfrac{1}{2}(\Delta-\sqrt{\Delta^2+4E_z^2p^2})$), and an excited Kramers pair $\left|u\pm\right\rangle= a_L\left|\Psi_{\pm 3/2}\right\rangle \mp ia_H\left|\Psi_{\mp1/2}\right\rangle$ (energy $\varepsilon_u=\tfrac{1}{2}(\Delta+\sqrt{\Delta^2+4E_z^2p^2})$).  Here, $a_L=\varepsilon_l/\sqrt{E_z^2p^2+\varepsilon_l^2}$ and $a_H=\sqrt{1-a_L^2}=E_zp/\sqrt{E_z^2p^2+\varepsilon_l^2}$. In the basis $\{\ket{l-},\ket{l+},\ket{u-},\ket{u+}\}$ Eq.~1 becomes
\begin{equation}
\bar{H}_{\rm op}=\left(
\begin{array}{cccc}
\varepsilon_{l} & \tfrac{1}{2}\lambda^*_{Zl} & \tfrac{1}{2}\lambda^*_{Zo} & 0 \\
\tfrac{1}{2}\lambda_{Zl} & \varepsilon_{l} & 0 & \tfrac{1}{2}\lambda_{Zo}\\
\tfrac{1}{2}\lambda_{Zo} & 0 & \varepsilon_{u} & \tfrac{1}{2}\lambda^*_{Zu} \\
0 & \tfrac{1}{2}\lambda^*_{Zo} & \tfrac{1}{2}\lambda_{Zu} & \varepsilon_{u} \\
\end{array}
\right).
\end{equation}
Here, the Zeeman terms $\lambda_{Zi}$ depend explicitly on $E_z$ due to the gate-induced mixing of $\ket{\Psi_{\pm 1/2}}$ and $\ket{\Psi_{\mp 3/2}}$.  We find $\lambda_{Zl}=2\varepsilon_Z(\sqrt{3}a_La_H-ia_L^2)$, $\lambda_{Zu}=2\varepsilon_Z(\sqrt{3}a_La_H-ia_H^2)$ and $\lambda_{Zo}=2\varepsilon_Z(-a_Ha_L+i\tfrac{\sqrt{3}}{2}a_L^2-i\tfrac{\sqrt{3}}{2}a_H^2)$. 

We perform a final rotation that exactly maps $\lambda_{Zl}$ and $\lambda_{Zu}$ to the diagonal, leaving $\lambda_{Zo}$ off-diagonal, defining a basis $\{\ket{-},\ket{+},\ket{e-},\ket{e+}\}$ (see Supplemental Material\cite{SMref}).  To zeroth order in $\lambda_{Zo}/(\varepsilon_u-\varepsilon_l)$, the splitting of the Kramers pair qubit states $\ket{+}$ and $\ket{-}$ is $\hbar\omega\approx|\lambda_{Zl}|$.  When mixed by the gate electric field, the spin 1/2 and spin 3/2 states with different Zeeman terms define a qubit $\ket{\pm}$ where $\hbar\omega$ is maximized (independent of electric fluctuations) $\mathbf{z}\delta E_z$ to first order when $\left|l\pm\right\rangle=\tfrac{\sqrt{3}}{2}\left|\Psi_{\pm1/2}\right\rangle \pm i (-\tfrac{1}{2})\left|\Psi_{\mp 3/2}\right\rangle$ (see Supplemental Material\cite{SMref}). As we will subsequently show, the qubit is also insensitive to in-plane electric noise $\delta E_{x,y}$, while a similar analysis yields another sweet spot at $E_z=0$.

Energy levels $\varepsilon_{\pm}=\varepsilon_l\pm\tfrac{1}{2}|\lambda_{Zl}|$ for the qubit are shown alongside excited levels $\varepsilon_{e\pm}=\varepsilon_u\pm\tfrac{1}{2}|\lambda_{Zu}|$ for $z_0=4.6$ nm ($6.9$ nm) in Fig.~\ref{fop}A (Fig.~\ref{fop}B).  Here, blue (red) hue denotes the amplitude of $a_L$ ($a_H$).  The qubit frequency is shown in Fig.~\ref{fop}C and Fig.~\ref{fop}D for approximate (black) and exact (green) solutions to $H_{\rm op}$, alongside the numerics (squares).  The maxima in $\hbar\omega$ in Fig.~\ref{fop}C (Fig.~\ref{fop}D) defines the sweet spot at $E_z=17$ MV/m ($14.8$ MV/m), for $|a_L|^2=3/4$, as expected.  We note that the approximate solution (Fig.~\ref{fop}C,D, black lines) captures the essential behaviour of the analytic model (Fig.~\ref{fop}C,D, green lines).  Corrections to Zeeman interactions from interface inversion asymmetry and cubic Land\'{e} g-factor, although included in the numerics (squares), have been neglected in the analytic model (green).  Note that the interface prevents ionization; although $E_z\sim15$ MV/m is much smaller than silicon's breakdown field, it well exceeds the ionization field of Si:B.\cite{Smit:2004fx}  

{\it In-plane electric fields: EDSR and noise immunity.}We express interactions with in-plane electric fields in the basis $\{\ket{-}, \ket{+}, \ket{e-}, \ket{e+}\}$, yielding
\small\begin{equation}
\tilde{H}=\left(\begin{array}{cccc}
\varepsilon_l-\tfrac{\hbar\omega}{2} & 0 & \alpha E_1 + \lambda_{Z_1} & \alpha E_2+\lambda_{Z_2}\\
0 & \varepsilon_l+\tfrac{\hbar\omega}{2} & \alpha E_2+  \lambda_{Z_2} & \alpha E_1+\lambda_{Z_1}\\
\alpha E_1^*+\lambda_{Z_1}^* & \alpha E^*_2+\lambda_{Z_2}^* & \varepsilon_u-\tfrac{|\lambda_{Zu}|}{2} & 0\\
\alpha E_2^*+\lambda_{Z_2}^* & \alpha E^*_1+\lambda_{Z_1}^* & 0 & \varepsilon_u+\tfrac{|\lambda_{Zu}|}{2}
\end{array}\right).
\end{equation}\normalsize
Here, $\ket{+}$ and $\ket{-}$ are our Kramers pair qubit states, $\lambda_{Z_1}\propto\lambda_{Zo}$ and $\lambda_{Z_2}\propto\lambda_{Zo}$ are Zeeman terms, and $E_{1,2}$ are interaction terms with in-plane electric fields, where $E_1=i(\sin\theta+\eta\cos\theta)E_x+i(\cos\theta+\eta\sin\theta)E_y$, $E_2=(-\cos\theta+\eta\sin\theta)E_x+(\sin\theta-\eta\cos\theta)E_y$, $\theta=\theta_u-\theta_l$, $\lambda_{Zi}=|E_{Zi}|\exp(i\theta_{i})$, and $\eta=p/\alpha$.  

The qubit Hamiltonian $H_{\rm qbt}=\hbar\omega\sigma_z + DE_{\parallel}\sigma_x$, where $\hbar\omega$ is the qubit frequency (Fig. 2C,D) and $D$ is the EDSR matrix element (Fig. 2E,F), is obtained by projecting the off-diagonal elements of $\tilde{H}$ to first order in $E_{x,y}$ using a Schrieffer-Wolff transformation.\cite{Schrieffer:1966hu,Winkler:2003dc}  Notably, qubit coherence is protected from in-plane electric noise since $\hbar\omega$ contains no terms to first order in $E_{x,y}$. EDSR drive comes from the transverse coupling $DE_{\parallel}\sigma_x$ in $H_{\rm qbt}$.  We obtain $D=\alpha|\lambda_{Zo}|(\varepsilon_l-\varepsilon_u)^{-1}(\alpha\cos(\theta_o-\theta_\parallel)+p\sin(\theta_o-\theta_\parallel))$, where $\mathbf{E}_{\parallel}=E_{\parallel}(\hat{\mathbf{x}}\cos\theta_\parallel+\hat{\mathbf{y}}\sin\theta_\parallel)$.  Interestingly, the small splitting $\varepsilon_u-\varepsilon_l$ essential for spin mixing at the sweet spot also causes strong EDSR, since $D\propto(\varepsilon_u-\varepsilon_l)^{-1}$. Note that the EDSR term is dominated by the inversion asymmetry quadrupolar SOC parameter $\alpha\approx 25$ D (Fig.~\ref{fop}F), since it is $100\times$ larger than the bare $T_d$ SOC parameter $p$.  

Importantly, $D$ can be maximized at the sweet spot by choosing the angle $\mathbf{E}_\parallel$ relative to $\mathbf{B}||\hat{\mathbf{y}}$ (see Fig.~2E,F). This yields fast gate times, but it also makes $D$, and therefore all timings based on EDSR, insensitive to fluctuations in electric field, protecting gate fidelity from noise at the Hamiltonian level. Since $\eta = p/\alpha\sim 0.01$ and $\theta_o=\pi/4$ at the sweet spot, $D$ is maximized with respect to $\theta_\parallel$ at $\theta_\parallel=-\pi/4\pm\pi/2$.  As shown for $z_0=4.6$ nm ($6.9$ nm) in Fig.~\ref{fop}E (Fig.~\ref{fop}F) $D$ is maximized with respect to $E_z$ for the same choice $\theta_\parallel$. This result can be easily obtained analytically, and holds for the analytic (green) and numerical (blue squares) solutions. 
 
{\it Qubit Operation}. The one-qubit and two-qubit gates employ EDSR-mediated interactions at the sweet spot, where coherence is protected from noise, and their times $\tau$ are minimized and also insensitive to electrical noise. EDSR driven $\pi$ rotations require $\tau_1=h/(2DE_{AC})=1$ ns ($0.2$ ns) for the $z_0=4.6$ nm (6.9 nm) deep acceptor, assuming a modest in-plane microwave field $E_{\rm AC}=500$ V/cm.  A $\pi/2$ (0) phase shift realizes a $\sigma_y$ ($\sigma_x$) gate, and $\sigma_z$ gates can be decomposed into a sequence of $\sigma_x$ and $\sigma_y$ gates. Readout can be accomplished by energy-dependent\cite{Pla:2012jj} or spin-dependent\cite{Petersson:2010ih} tunneling, or dispersive readout in cQED.\cite{Blais:2004kn} Initialization can be achieved by projective readout followed by spin rotation.  

Two-qubit entanglement can be achieved via long-ranged Coulomb interactions, owing to spin-dependent electric dipole-dipole interactions.\cite{Golovach:2006hx,Flindt:2006dn,Trif:2007fx} Their strength is given by $J_{dd}=(\mathbf{v}_1 \cdot \mathbf{v}_2 R^2 - 3(\mathbf{v}_1 \cdot \mathbf{R})(\mathbf{v}_2\cdot\mathbf{R}))/4\pi\epsilon R^5$, where $\mathbf{R}$ is the inter-qubit displacement and $\mathbf{v}_i$ is a spin-dependent charge dipole of qubit $i$, which has the same magnitude as the EDSR matrix element. For a $20$ nm distance with negligible tunnel coupling, we obtain a $\sqrt{\textrm{SWAP}}$ time of $\tau_{2dd}=h/4J_{dd}\approx 2$ ns with $J_{dd}\approx D^2/4\pi\epsilon R^3$.  The $10^2$ times enhancement of EDSR from the interface reduces $\tau_{2dd}$ by $10^4$ relative to acceptors in bulk silicon, and $10^5$ relative to bare magnetic dipole-dipole coupling.  Entanglement by Heisenberg exchange is also possible and exchange is hydrogenic when $\Delta$ exceeds $J$.\cite{Salfi:2016uma}  We note that the advantage that holes do not have valley degrees of freedom\cite{Salfi:2014kaa} which may complicate Heisenberg exchange for electrons in Si.\cite{Koiller:2001gw}  

{\it Circuit QED}. Coplanar superconducting microwave cavities could be used to implement cQED including two-qubit gates, dispersive single-spin readout, and strong Jaynes-Cummings coupling on resonance with the cavity.\cite{Blais:2004kn,Wallraff:2004dy,Xiang:2013hm}  We assume a coplanar waveguide resonator operating at $B=0.5$ T ($f=15$ GHz) and a vacuum electric field $E_0\approx 50$ V/m.  This can be obtained using a tapered resonator gap, or a superconducting nanowire resonator.\cite{Samkharadze:2015tw}  At the sweet spot for $z_0=4.6$ nm ($6.9$ nm), the vacuum Rabi coupling is $g_c=eDE_0=2$ neV ($10$ neV).    

For cavity mediated non-demolition readout and qubit coupling, we detune the qubit from the cavity by $\Delta = 4g_c$.\cite{Kloeffel:2013iv}  Here, the spin state shifts the cavity resonance by $\Delta f = g_c^2/\Delta=0.25$ MHz (1.25 MHz) for $z_0=4.6$ nm ($6.9$ nm).  The two-qubit $\sqrt{\textrm{SWAP}}$ time is $\tau_{2c}=h/4J_{c}=200$ ns for $z_0=6.9$ nm, determined by the effective spin-spin interaction\cite{Kloeffel:2013iv} $J_{c}=2g_c^2/\Delta=2.5$ MHz.  Operating at zero detuning, spin/photon Rabi oscillations require $\hbar\pi/g_c=1$ $\mu$s ($200$ ns).  Assuming $Q=10^5$ at $B_0=0.5$ T in state-of-the-art superconducting cavities\cite{Graaf:2012fl,Samkharadze:2015tw} $g_c\kappa = 6.7$ (33) Rabi cycles can be obtained for $z_0=4.6$ nm ($6.9$ nm), where $\kappa=f/Q$ is the cavity loss rate. 

{\it Relaxation and Dephasing}. We consider spin-lattice (phonon) relaxation and dephasing from a host of electrical noise sources, and compare them to gate times.  Since silicon is not piezoelectric, spin-lattice relaxation occurs only via the deformation potential\cite{Ehrenreich:1956bq,Srivastava:1990tk}.  For temperatures $T\ll \hbar\omega/k_B$, the spin relaxation time derived in the Supplemental Material\cite{SMref} follows $T_1^{-1}=(\hbar\omega)^3(C_d/20\rho\pi\hbar^4)(|\lambda_{Zo}|/\Delta)^2$, where $|\lambda_{Zo}|/\Delta=\hbar\omega/4pE_z$ at the sweet spot, and $C_d=4.9\times10^{-20}$ (eV)$^2$(s/m)$^5$.  We obtain $T_1=20$ $\mu$s (5 $\mu$s) for $z_0=4.6$ nm (6.9 nm) at $B_0=0.5$ T that are $100$ times longer \textit{c.f.} bulk unstrained silicon at $B=0.5$ T.\cite{Bir:1963iy,Ruskov:2013kq}  

Random fluctuations in qubit splitting $\hbar\delta\omega(t)$ dephase the qubit.  The dephasing rate from random telegraph signal (RTS) in charge trap occupation is $(T_2^*)^{-1}=(\delta\omega)^2\tau_S/2$, where $\delta\hbar\omega$ is qubit frequency shift, and $\tau_S$ is the average switching time.\cite{Bermeister:2014jb}  We take $\tau_S=10^3\tau_1$ as the worst case, since slower fluctuations can be suppressed by dynamical decoupling.  Assuming a trap $50$ nm away, we find $\delta E \sim2,000$ V/m and a large window of $200,000$ V/m ($20,000$ V/m) of gate space where $T_2^*>2T_1$ at the sweet spot, for $z_0=4.6$ nm ($6.9$ nm). In comparison, the same analysis gives $T_2^* \sim 0.1$ ns for acceptor-based charge qubits with similar gate times.  It is remarkable that in comparison, electrical noise has virtually no effect on coherence in our spin-orbit qubit, illustrating the advantages of inversion asymmetry and our spin-orbit qubit's sweet spot. We also find that dephasing from Johnson-limited gate voltage noise, and from two-level (tunneling) systems (TLS), are $\sim 10^7$ and $\sim 10^4$ times weaker, respectively, compared with RTS.\cite{Bermeister:2014jb} There are only a few spin resonance experiments on acceptors\cite{Hensel:1963dd,Neubrand:1978je,Neubrand:1978jz,Kopf:1992gj,Tezuka:2010cl,Stegner:2010is,Song:2011jn}, none of which feature strain and an interface.\cite{Mol:2015im}  We expect hyperfine-induced decoherence in $^{\rm nat}$Si to be weak since it has only 4.7 \% of spin-bearing isotopes and hyperfine interactions are weaker for holes than electrons.\cite{Chekhovich:2012ks}  Meanwhile, $^{28}$Si enrichment could be used to virtually eliminate the nuclear bath.\cite{Tyryshkin:2011fi}

The insensitivity to Johnson noise and tunneling TLS means spin-lattice $T_1$ limits coherence for few (or slow enough) traps at Si/SiO$_2$ interfaces.  For $B=0.5$ T, $r_1>10^4$ single qubit gates, $r_{2dd}>10^3$ dipole-dipole two-qubit gates, and $r_{2c}\approx 25$ cavity-mediated two-qubit gates can be achieved in a $T_1$ limited coherence time.  Therefore while $T_1$ is short compared to donors, many gate operations can performed.  Since $T_1\propto \omega^{-5}$, choosing $B=0.25$ T increases all ratios favourably to $r_1>10^5$, $r_{2dd}\approx 10^4$, and $r_{2c}\approx 50$.  Since $T_1$ is much longer at the $E_z=0$ sweet spot, adiabatically sweeping to $E_z=0$ opens a pathway for a long-lived quantum memory.

{\it Conclusions}. The proposed single-acceptor spin-orbit qubit exploits the tunability of the $J=3/2$ manifold of acceptors and the associated quadrupolar SOC arising from the ion and interface potential, providing for (i) fast one-qubit and long-ranged two-qubit gates (ii) at a sweet spot where the qubit phase and all gate timings are insensitive to electrical fluctuations, (iii) avoiding entirely the need for exchange interactions, (iv) in an industrially relevant silicon platform. $10^5$ single-qubit and $10^4$ two-qubit gates could be possible in the qubit coherence time.  Using cQED, dispersive single-spin readout, cavity-mediated spin-spin entanglement, and Jaynes-Cummings spin-photon entanglement are possible. 

\begin{acknowledgements}We thank R. Winkler, U. Zuelicke, M. Tong, and T. Kobayashi for helpful discussions. JS, JAM and SR acknowledge funding from the ARC Centre of Excellence for Quantum Computation and Communication Technology (CE110001027), and in part by the US Army Research Office (W911NF-08-1-0527).  DC acknowledges funding through the ARC Discovery Project scheme.\end{acknowledgements}

\section{Appendix}
\renewcommand{\thefigure}{A\arabic{figure}}
\renewcommand{\thetable}{A\arabic{table}}
\renewcommand{\theequation}{A\arabic{equation}}
\setcounter{figure}{0}
\setcounter{equation}{0}
\setcounter{table}{0}
\section{Interactions with magnetic fields and strain}
\label{app:zeeman}

Interactions of acceptor-bound holes with magnetic fields and strain are known for acceptor dopants in bulk silicon\cite{Bir:1963iy,Bir:1963cu}. In the $\left|\Psi_{m_J}\right\rangle$ subspace, interactions with magnetic fields $\mathbf{B}=\hat{\mathbf{x}}B_x+\hat{\mathbf{y}}B_y+\hat{\mathbf{z}}B_z$ are represented by the Hamiltonian
\begin{equation}
H_Z=\mu_B(g_1(J_xB_x+\textrm{c.p.})+g_2(J_x^3B_x+\textrm{c.p})).
\end{equation}
Here, $J_\alpha$ are $J=3/2$ matrices, c.p. refers to cyclic permutations, $g_1$ and $g_2$ are the linear and cubic Land\'{e} g-factors, and $\mu_B$ is the Bohr magneton.  Interactions with strain $\epsilon_{ij}$ are represented by the Hamiltonian
\begin{align}
H_\epsilon=&a'\textrm{Tr}[\epsilon] + b'((J_x^2-\tfrac{5}{4}I)\epsilon_{xx}+\textrm{c.p.})\nonumber\\+&(2d'/\sqrt{3})(\{J_x,J_y\}\epsilon_{xy}+\textrm{c.p.}),
\end{align}
where $\{J_x,J_y\}=\tfrac{1}{2}(J_xJ_y+J_yJ_x)$, $a'$, $b'$ and $d'$ are Bir-Pikus deformation potentials\cite{Bir:1963iy,Bir:1963cu}.

\section{Acceptor states in spherical spin-3/2 basis}
\label{app:states}
In the spherical spin-3/2 basis $\left|L,J;F,m_F\right\rangle$, where $\mathbf{F}=\mathbf{L}+\mathbf{J}$ is an effective total angular momentum\cite{Baldereschi:1973im}, acceptor eigenstates take the form
\begin{align}
\left|\Psi_{m_J}\right\rangle=f_0(r)&\left|L=0,J=\tfrac{3}{2};F=\tfrac{3}{2},m_F\right\rangle\nonumber\\
+g_0(r)&\left|L=2,J=\tfrac{3}{2};F=\tfrac{3}{2},m_F\right\rangle,
\end{align}
where $J=\tfrac{3}{2}$ is an effective total spin, and the spin-3/2 spin-orbit interaction has coupled states with $\Delta L=0,\pm2$.  The $f_0(r)$ and $g_0(r)$ are radial envelope wavefunctions for envelope function spherical harmonics with $L=0$ and $L=2$, respectively.  The $\left|L,J;F,m_F\right\rangle$ are found using the Clebsch-Gordan coefficients.  

States outside the $4\times4$ subspace, given in ref.~\cite{Baldereschi:1973im}, contribute to envelope function asymmetry to realize the $b$ and $d$ terms in $H_E$ from the main text.  The Schrieffer-Wolff calculation mixing in these higher excited states will be presented in a future paper.

\section{Analytic model for low energy states}
\label{app:unitary}
The low-energy holes are described in the $\left|\Psi_{m_J}\right\rangle$ basis by
\begin{equation}
H_{\rm op}=\left(
\begin{array}{cccc}
0 & -i\tfrac{\sqrt{3}}{2}\varepsilon_Z & -ipE_z & 0 \\
i\tfrac{\sqrt{3}}{2}\varepsilon_Z & \Delta(E_z) & -i\varepsilon_Z & -ipE_z \\
ipE_z & i\varepsilon_Z & \Delta(E_z) & -i\tfrac{\sqrt{3}}{2}\varepsilon_Z \\
0 & ipE_z & i\tfrac{\sqrt{3}}{2}\varepsilon_Z & 0\\
\end{array}
\right)
\end{equation}
for $m_J=+\tfrac{3}{2},+\tfrac{1}{2},-\tfrac{1}{2},-\tfrac{3}{2}$, where $\varepsilon_Z=g_1\mu_BB$, and  $g_1=1.07$ for B:Si\cite{Neubrand:1978je,Kopf:1992gj}. The unitary transform $U_0$ that diagonalizes $H_{\rm op}(\varepsilon_Z=0)$ is
\begin{equation}
U_0=\left(
\begin{array}{cccc}
-ia_H & 0    & 0    & a_L\\
0     & a_L  & ia_H & 0\\
a_L   & 0    & 0    & -ia_H\\
0     & ia_H & a_L  & 0
\end{array}\right),
\end{equation}
in the basis $\{\left|\Psi_{+3/2}\right\rangle,\left|\Psi_{+1/2}\right\rangle,\left|\Psi_{-1/2}\right\rangle,\left|\Psi_{-3/2}\right\rangle\}$.  Here, $a_L=\varepsilon_l/\sqrt{E_z^2p^2+\varepsilon_l^2}$ and $a_H=\sqrt{1-a_L^2}=E_zp/\sqrt{E_z^2p^2+\varepsilon_l^2}$.
Applying this transformation we obtain an extended qubit Hamiltonian for the spin qubit states and nearest excited spin states
\begin{equation}
\bar{H}_{\rm op}=\left(
\begin{array}{cccc}
\varepsilon_{l} & \tfrac{1}{2}\lambda^*_{Zl} & \tfrac{1}{2}\lambda^*_{Zo} & 0 \\
\tfrac{1}{2}\lambda_{Zl} & \varepsilon_{l} & 0 & \tfrac{1}{2}\lambda_{Zo}\\
\tfrac{1}{2}\lambda_{Zo} & 0 & \varepsilon_{u} & \tfrac{1}{2}\lambda^*_{Zu} \\
0 & \tfrac{1}{2}\lambda^*_{Zo} & \tfrac{1}{2}\lambda_{Zu} & \varepsilon_{u} \\
\end{array}
\right).
\end{equation}
in the basis $\{\left|l-\right\rangle,\left|l+\right\rangle,\left|u-\right\rangle\},\left|u+\right\rangle\}$.  Here, the effective Zeeman interactions $\lambda_{Zi}$ appear on the off-diagonal.  The effective Zeeman interactions depend explicitly on the gate electric field due to the $pE_z$-induced mixing of $\left|\Psi_{\pm 1/2}\right\rangle$ and $\left|\Psi_{\mp 3/2}\right\rangle$.  These off-diagonal Zeeman interactions are
\begin{align}
\lambda_{Zl}&=2\varepsilon_Z(\sqrt{3}a_La_H-ia_L^2),\\
\lambda_{Zu}&=2\varepsilon_Z(\sqrt{3}a_La_H-ia_H^2), \textrm{ and}\\
\lambda_{Zo}&=2\varepsilon_Z(-a_Ha_L+i\sqrt{3}a_L^2/2-i\sqrt{3}a_H^2/2).
\end{align}
The upper and lower $2\times2$ blocks of $\bar{H}_{\rm op}$ are diagonalized by
\begin{equation}
U_{Z0}=2^{-\tfrac{1}{2}}\left(
\begin{array}{cccc}
+e^{-i\theta_l/2} & +e^{-i\theta_l/2} & 0 & 0 \\
-e^{+i\theta_l/2} & +e^{+i\theta_l/2} & 0 & 0 \\
0 & 0 & +e^{-i\theta_u/2} & +e^{-i\theta_u/2} \\
0 & 0 & -e^{+i\theta_u/2} & +e^{+i\theta_u/2}\end{array}
\right),
\end{equation}
in the basis $\{\left|l-\right\rangle,\left|l+\right\rangle,\left|u-\right\rangle,\left|u+\right\rangle\}$ where,
\begin{align}
\theta_l&=\textrm{arctan}(\sqrt{3}a_Ha_L,-a_L^2),\textrm{ and}\\
\theta_u&=\textrm{arctan}(\sqrt{3}a_La_H,-a_H^2).
\end{align}
Applying this transformation we obtain the following extended Hamiltonian 
\begin{equation}
\tilde{H}_{\rm op}=\left(
\begin{array}{cccc}
\varepsilon_l-\tfrac{|\lambda_{Zl}|}{2} & 0 & \lambda_{Z1} & \lambda_{Z2} \\
0 & \varepsilon_l+\tfrac{|\lambda_{Zl}|}{2} & \lambda_{Z2} & \lambda_{Z1} \\
\lambda_{Z1}^* & \lambda_{Z2}^* & \varepsilon_u-\tfrac{|\lambda_{Zu}|}{2} & 0 \\
\lambda_{Z2}^* & \lambda_{Z1}^* & 0 & \varepsilon_u+\tfrac{|\lambda_{Zu}|}{2}
\end{array}
\right)
\label{eq:tHop}
\end{equation}
in the basis $\{\left|-\right\rangle,\left|+\right\rangle,\left|e-\right\rangle,\left|e+\right\rangle\}$, where,
\begin{align}
\lambda_{Z1}&=\tfrac{1}{2}|\lambda_{Zo}|\cos(\theta_l/2-\theta_u/2-\theta_o),\\
\lambda_{Z2}&=\tfrac{1}{2}|\lambda_{Zo}|i\sin(\theta_l/2-\theta_u/2-\theta_o),\textrm{ and}\\
\theta_o&=\textrm{arctan}(\tfrac{\sqrt{3}}{2}(a_L^2-a_H^2),a_Ha_L).
\end{align}
Our approximate qubit model from the main text takes $\lambda_{Zo}/(\varepsilon_u-\varepsilon_l)$ to zeroth order to give qubit states $\left|-\right\rangle$ and $\left|+\right\rangle$. In this approximation we obtain a qubit frequency
\begin{equation}
\hbar\omega=|\lambda_{Zl}|=2\varepsilon_Z\sqrt{3a_L^2a_H^2+a_L^4}.
\end{equation}
A sweet spot occurs when the qubit frequency is insensitive to small fluctuations in electric fields.  For an acceptor experiencing no static external applied field along $x$ and $y$ directions, this occurs for the roots of 
\begin{equation}
\partial\hbar\omega/\partial E_z=2\varepsilon_Z(\partial a_L/\partial E_z)(3-4a_L^2)/\sqrt{3-2a_L^2}.
\end{equation}
One root (sweet spot) is $(a_L,a_H)=(-\tfrac{\sqrt{3}}{2},\tfrac{1}{2})$.  Substituting $\varepsilon_l$ and $\Delta(E_z)$, we obtain an equivalent condition
\begin{equation}
\Delta_G(E_z) + \Delta_{\rm if} + 2p/\sqrt{3}E_z  = \Delta_{\epsilon}
\end{equation}
for this sweet spot.  Another root (sweet spot) occurs at the roots of
\begin{equation}
\partial a_L/\partial E_z=E_z^2p^2(\partial \varepsilon_l/\partial E_z)(\varepsilon_l^2+E_z^2p^2)^{-3/2},
\end{equation}
The sweet spot associated with the above root occurs at $E_z=0$.  The lowest order correction to $\hbar\omega$ due to coupling to levels $\left|e\pm\right\rangle$ is easily obtained from 2$^{\textrm{nd}}$-order perturbation theory,
\begin{equation}
\delta\hbar\omega=-\frac{1}{4}\Big(\frac{|\lambda_{Zo}|}{\varepsilon_l-\varepsilon_u}\Big)^2(|\lambda_{Zl}|-|\lambda_{Zu}|\cos(2\theta_o-\theta_l+\theta_u)).
\end{equation}
The exact solution in (main text, Fig.~2C,D) shows that all higher order corrections (including $\hbar\omega^{(2)}$) to the approximate solution presented in the main text do not qualitatively modify the qubit frequency.

\section{Electric dipole spin resonance}
\label{app:EDSR}
The total interaction with in-plane electric fields $E_x$ and $E_y$ described by $H_{E}$ and $H_{E,\rm ion}$ is 
\begin{align}
H_{E_\parallel}=&\alpha(E_z)\left(\begin{array}{cccc}
0 & E_- & 0 & 0\\
E_+ & 0 & 0 & 0\\
0 & 0 & 0 & -E_-\\
0 & 0 & -E_+ & 0
\end{array}\right)\nonumber\\
+&p\left(\begin{array}{cccc}
0 & -iE_+ & 0 & 0\\
iE_- & 0 & 0 & 0\\
0 & 0 & 0 & +iE_+\\
0 & 0 & -iE_- & 0
\end{array}\right),
\end{align}
in the basis $\{\left|\Psi_{+3/2}\right\rangle,\left|\Psi_{+1/2}\right\rangle,\left|\Psi_{-1/2}\right\rangle,\left|\Psi_{-3/2}\right\rangle\}$, where $E_+=E_x+iE_y$ and $E_-=E_x-iE_y$.  The first matrix is the coupling due to the broken inversion symmetry of the interface and gate field, along the $z$ direction, while the second matrix describes the interaction due to the $T_d$ symmetry of the local field of the ion. Rotated into the qubit basis using $U_{t0}=U_{Zo}U_0$, we obtain
\begin{equation}
\tilde{H}=\left(\begin{array}{cccc}
\varepsilon_l-\tfrac{|\lambda_{Zl}|}{2} & 0 & \alpha E_1 + Z_1 & \alpha E_2+Z_2\\
0 & \varepsilon_l+\tfrac{|\lambda_{Zl}|}{2} & \alpha E_2+  Z_2 & \alpha E_1+Z_1\\
\alpha E_1^*+Z_1^* & \alpha E^*_2+Z^*_2 & \varepsilon_u-\tfrac{|\lambda_{Zu}|}{2} & 0\\
\alpha E_2^*+Z_2^* & \alpha E^*_1+Z^*_1 & 0 & \varepsilon_u+\tfrac{|\lambda_{Zu}|}{2}
\end{array}\right),
\label{eq:HEsup}
\end{equation}
in the basis $\{\left|-\right\rangle,\left|+\right\rangle,\left|e-\right\rangle,\left|e+\right\rangle\}$, where, $E_1=i(\sin\theta+\eta\cos\theta)E_x+i(\cos\theta+\eta\sin\theta)E_y$, $E_2=(-\cos\theta+\eta\sin\theta)E_x+(\sin\theta-\eta\cos\theta)E_y$, $\theta=\theta_u-\theta_l$, and $\theta_i=\textrm{arg}(\lambda_{Zi})$, and $\eta=p/\alpha$.  

EDSR coupling $DE_{\parallel}\sigma_x$ in the qubit subspace arises from oscillating in-plane electric fields $\mathbf{E}_{\parallel}=E_{\parallel}(\hat{\mathbf{x}}\cos\theta_\parallel+\hat{\mathbf{y}}\sin\theta_\parallel)$.  We obtain the effective EDSR interactions $DE_\parallel\sigma_x$ in the qubit basis using a Schrieffer-Wolff transformation.  Expanding in $\eta_l=|\lambda_{Zl}|/(\varepsilon_l-\varepsilon_u)$ and $\eta_u=|\lambda_{Zu}|/(\varepsilon_l-\varepsilon_u)$ and grouping terms according to powers in the electric fields $E_1$ and $E_2$, we obtain
\begin{align}
D=&\frac{2|\lambda_{Zu}|\textrm{Re}(\lambda_{Z1}^*\lambda_{Z2})}{(\varepsilon_l-\varepsilon_u)^2}\nonumber\\
+&\alpha\frac{2\textrm{Re}(E^*_2\lambda_{Z1} + E^*_1\lambda_{Z2})}{\varepsilon_l-\varepsilon_u}\nonumber\\
+&\alpha^2\frac{2|\lambda_{Zu}|\textrm{Re}(E^*_1E_2)}{(\varepsilon_l-\varepsilon_u)^2}.
\end{align}
The term proportional to $E^*_2\lambda_{Z1} + E^*_1\lambda_{Z2}$ is linear in the electric field and defines the electric dipole spin resonance term. Substituting $\lambda_{Z1}$, $\lambda_{Z2}$, $E_1$ and $E_2$ we obtain the EDSR matrix element 
\begin{equation}
D=\alpha\frac{|\lambda_{Zo}|}{\varepsilon_l-\varepsilon_u}(\cos(\theta_o-\theta_\parallel)+\eta\sin(\theta_o-\theta_\parallel)).
\end{equation}
\section{Spin-dependent dipole-dipole interaction}

Because of the spin-orbit interaction, the electric dipole moment in each acceptor couples to spin.  As a result, two qubits interacting only via mutual Coulomb repulsion experience a spin-dependent interaction resembling a magnetic dipole-dipole interaction. Here we determine this interaction in the coupled-qubit subspace $\ket{--}, \ket{-+}, \ket{+-}, \ket{++}$.  For the total Hamiltonian we have $H^{\Sigma}=H_{\rm op}^1+H_{\rm op}^2+V^{12}$, where $H^i_{\rm op}$ is the single-acceptor Hamiltonian for qubit $i=1,2$, and $V_{12}(\mathbf{r}_1-\mathbf{r}_2)=e^2/4\pi\epsilon|\mathbf{r}_1-\mathbf{r}_2|$ is the electrostatic interaction between the qubits. 

We work in the tensor $16\times16$ tensor product subspace of the two qubits $\ket{mn}=\ket{m^1}\otimes\ket{n^2}$ where $m\in\{-,+,u-,u+\}$ and $n\in\{-,+,u-,u+\}$, explicitly ignoring anti-symmetrization, \textit{i.e.}, assuming spatial overlaps are negligible. Without the Coulomb interaction, the Hamiltonian is
\begin{align}
\braket{mn|H_{\rm op}^1+H_{\rm op}^2|m'n'}=&\braket{m|H_{\rm op}^1|m'}\delta_{nn'}\nonumber\\
+&\braket{n|H_{\rm op}^2|n'}\delta_{mm'},
\end{align}
where $H^i_{\rm op}$ is given by Equation (\ref{eq:tHop}). Meanwhile, in the direct product subspace the two-qubit Coulomb interaction is
\begin{align}
&\braket{mn|V^{12}|m'n'}\nonumber\\
=&\int dr_1^3 dr_2^3 \frac{e^2\Psi_m^\dagger(\mathbf{r}_1)\Psi_{n}^\dagger(\mathbf{r}_2)\Psi_{m'}(\mathbf{r}_1)\Psi_{n'}(\mathbf{r}_2)}{4\pi\epsilon|\mathbf{r}_1-\mathbf{r}_2|}\label{eq:V12h}
\end{align}
When the separation $\mathbf{R}_{12}$ between the acceptors is large compared to dipole moments $\left\langle\delta\mathbf{r}_i\right\rangle$ of the system, we may use the multi-pole expansion of the Coulomb interaction in Equation (\ref{eq:V12h}).  The lowest-order non-zero term is
\begin{align}
\braket{mn|V^{12}|m'n'}=\frac{e^2}{4\pi\epsilon R_{12}^5}\Big[R_{12}^2\braket{\delta\mathbf{r}_1}_{nn'}\cdot\braket{\delta\mathbf{r}_2}_{mm'}&\nonumber\\-3(\braket{\delta\mathbf{r}_1}_{nn'}\cdot\mathbf{R}_{12})(\braket{\delta\mathbf{r}_2}_{mm'}\cdot\mathbf{R}_{12})&\Big]\label{eq:d12d}
\end{align}
where 
\begin{equation}
\braket{\delta\mathbf{r}_i}_{nn'}=\int dr^3_i (\mathbf{r}_i-\mathbf{R}_i) \Psi_n^\dagger(\mathbf{r}_i)\Psi_{n'}(\mathbf{r}_i),
\end{equation}
$\mathbf{R}_i$ is the position of qubit ion $i$, and $\mathbf{R}_{12}=\mathbf{R}_1-\mathbf{R}_2$. The Coulomb interaction is now a product of single-hole dipole matrix elements known in the basis $\{-,+,u-,u+\}$ from Equation (\ref{eq:HEsup}). The total Hamiltonian is
\begin{align}
\braket{mn|H_{\Sigma}|m'n'}=&\braket{m|H_{\rm op}^1|m'}\delta_{mm'}+\braket{n|H_{\rm op}^2|n'}\delta_{nn'}\nonumber\\
+&\braket{mn|V^{12}|m'n'}
\end{align}
The combined effect of the Coulomb and Zeeman interactions can be projected into the coupled-qubit subspace $\ket{--}, \ket{-+}, \ket{+-}, \ket{++}$ using a Schrieffer-Wolff transformation.  Working out the effective interaction, to second order in off-diagonal terms and zeroth order in $|\lambda_{Zu}|$ and $|\lambda_{Zl}|$, gives a spin-independent shift to all levels which does not influence the qubit physics.  A transformation to third order in off-diagonal terms in the full $16\times16$ space finds the spin-spin interaction of interest,
\begin{equation}
H_{\rm dd}=J_{xx}\left(
\begin{array}{cccc}
0 & 0 & 0 & 1\\
0 & 0 & 1 & 0\\
0 & 1 & 0 & 0\\
1 & 0 & 0 & 0
\end{array}\right),
\label{eq:Jdd}
\end{equation}
which is an Ising type spin-spin interaction.  For convenience this can be re-written as
\begin{align}
H_{dd}=&J_{xx}\sigma_{1x}\sigma_{2x}\\=&J_{xx}(\sigma_{1+}+\sigma_{1-})(\sigma_{2+}+\sigma_{2-}),
\end{align}
where $\sigma_{\pm}=\sigma_{jx}\pm i\sigma_{jy}$ is the raising/lowering operator for qubit $j$.  For $\mathbf{R}_{12}=\hat{\mathbf{x}}R_{12}$, we have contributions $J_{xx,x}$ and $J_{xx,y}$ from the $x$ and $y$ oriented dipoles, respectively such that the total coupling is $J_{xx}=-2J_{xx,x}+J_{xx,y}$, where
\begin{equation}
J_{xx,\mu}=\frac{4(q_{2\mu}|\lambda_{Z1}|+q_{1\mu}|\lambda_{Z2}|)^2}{4\pi\epsilon R_{12}^3(\varepsilon_u-\varepsilon_l)^2},
\end{equation}
$q_{1x}=\alpha\sin(\theta)$, $q_{2x}=-\alpha\cos(\theta)$, $q_{1y}=\alpha\cos(\theta)$, and $q_{2y}=\alpha\sin(\theta)$.  Substituting $|\lambda_{Z1}|=\frac{1}{2}|\lambda_{Zo}|\cos(-\theta/2-\theta_o)$, $|\lambda_{Z2}|=\frac{1}{2}|\lambda_{Zo}|\sin(-\theta/2-\theta_o)$, we obtain
\begin{align}
J_{xx,x}&=\frac{\alpha^2|\lambda_{Zo}|^2\cos^2\theta_o}{4\pi\epsilon R^3(\varepsilon_u-\varepsilon_l)^2},\textrm{ and}\\
J_{xx,y}&=\frac{\alpha^2|\lambda_{Zo}|^2\sin^2\theta_o}{4\pi\epsilon R^3(\varepsilon_u-\varepsilon_l)^2},
\end{align}
such that 
\begin{align}
J_{xx}&=\Big(\frac{\alpha|\lambda_{Zo}|}{\varepsilon_u-\varepsilon_l}\Big)^2\frac{(-2\cos^2\theta_o+\sin^2\theta_o)}{4\pi\epsilon R^3}\\
&\approx\frac{D^2}{4\pi\epsilon R^3}.
\end{align}

\section{Phonon-induced spin relaxation}
\label{app:T1}
The relaxation from $\left|n'\right\rangle$ to $\left|n\right\rangle$ via emission of a phonon with energy $\hbar\omega_{qs}=\hbar v_s q_s$ can be determined from Fermi's golden rule,
\begin{align}
\frac{1}{T_{n\rightarrow n'}}=\frac{2\pi}{\hbar}\sum_{i,j,s,\mathbf{q}_s}&|\left\langle n',n_{\mathbf{q}}+1|H_{\epsilon ij s}|n,n_\mathbf{q}\right\rangle|^2\nonumber\\
\times&\delta(E_n-E_n'-\hbar\omega_{qs})
\end{align}
where $s=\ell,t_1,t_2$ are the phonon polarizations, $\mathbf{q}_s$ is the phonon wavevector, $\sum_{i,j}H_{\epsilon ij s}=\sum_{i,j}D_{ij}\epsilon_{ijs}$ is the electron-phonon interaction, and $n_{\mathbf{q}}$ is the phonon population. The deformation potential matrices $D_{ij}$ are determined from the Bir-Pikus Hamiltonian
\begin{align}
\sum D_{ij}\epsilon_{ijs}=&a'(\epsilon_{xxs}+\textrm{c.p.})\nonumber\\+&b'[(J_x^2-\tfrac{5}{4}I)\epsilon_{xxs}+\textrm{c.p.}]\nonumber\\+&(2d'/\sqrt{3})[\{J_x,J_y\}\epsilon_{xys}+\textrm{c.p.}]
\end{align} 
where $a'$, $b'$, and $d'$ are Bir-Pikus deformation potentials\cite{Bir:1963iy,Bir:1963cu} and the strain $\epsilon_{ijs}=\tfrac{1}{2}(\partial\delta R_{is}/\partial r_j+\partial\delta R_{js}/\partial r_i)$ of the phonon polarization $s$ is determined by the displacement\cite{Srivastava:1990tk}
\begin{equation}
\delta\mathbf{R}_s=(-i)\sqrt{\frac{\hbar}{2NV\rho\omega_{qs}}}\hat{\mathbf{e}}_{qs}(a^\dagger_{\mathbf{q}s}+a_{\mathbf{q}s})\exp(i\mathbf{q}_s\cdot\mathbf{r}),
\end{equation}
where $\hat{\mathbf{e}}_{qs}$ is the normalized phonon polarization vector\cite{Ehrenreich:1956bq}, $N$ is the number of unit cells, $V$ is the unit cell volume, $NV=L^3$ is the crystal volume, $\rho$ is the mass density, and $a^\dagger_{\mathbf{q}s}$ ($a_{\mathbf{q}s}$) creates (destroys) a phonon of wavevector $\mathbf{q}_s$ and polarization $s$. For $\left\langle n'|D_{ij}\exp(i\mathbf{q}\cdot\mathbf{r})|n\right\rangle$ we use the dipole approximation $\left\langle n'|D_{ij}(1+i\mathbf{q}\cdot\mathbf{r}+\dots)|n\right\rangle\approx\left\langle n'|D_{ij}|n\right\rangle$, which is appropriate since $qa\sim10^{-2}$ where $q$ is the the phonon wavevector and $a\sim1$ nm is the Bohr radius.   At low temperatures $T \ll \hbar\omega/k \approx 0.7$ K we obtain
\begin{align}
\frac{1}{T_1}=\frac{(\hbar\omega)^3}{20\hbar^4\pi\rho}\Big[&\sum_i |\left\langle n' | D_{ii}| n \right\rangle|^2 \Big(\frac{2}{v_l^5}+\frac{4}{3v_t^5}\Big)\nonumber\\
+&\sum_{i\neq j}|\left\langle n' | D_{ij}| n \right\rangle|^2\Big(\frac{2}{3v_l^5}+\frac{1}{v_t^5}\Big)\Big]
\end{align}
where $v_l$ and $v_t$ are the longitudinal and transverse sound velocities. For our qubit, the relaxation from $\left|+\right\rangle$ to $\left|-\right\rangle$ can be evaluated using a Schrieffer-Wolff transformation of the non-diagonal elements of $\tilde{H}=\tilde{H}_{\rm op}+\tilde{H}_{\epsilon i j s}$, where $\tilde{H}_{\epsilon i j s}=U^\dagger_{t0}(\sum_{i,j,s} H_{\epsilon i j s}\epsilon_{ijs})U_{t0}$. We determine the coupling to lowest order in $\tilde{H}_{Zo}$ and $\tilde{H}_{\epsilon i j s}$ while treating $H_{E,\rm ion}$, $H_{\rm if}$ and $H_{E}$ exactly within the $4\times4$ subspace. We obtain 
\begin{align}
\frac{1}{T_1}=\frac{(\hbar\omega)^3}{20\hbar^4\pi\rho}\Big(\frac{\hbar\omega}{4pE_z}\Big)^2\Big[&\frac{9}{32}b'^2\Big(\frac{2}{v_l^5}+\frac{4}{3v_t^5}\Big)\nonumber\\+&\frac{5}{16}d'^2\Big(\frac{2}{3v_l^5}+\frac{1}{v_t^5}\Big)\Big]
\end{align}
at the sweet spot. For comparison, we also evaluated the phonon-mediated transition rate from the first excited level $\left|-\frac{1}{2}\right\rangle$ to the ground state $\left|-\frac{3}{2}\right\rangle$ in bulk unstrained silicon, which is allowed even in zero magnetic field, since $\left|-\frac{1}{2}\right\rangle$ and $\left|-\frac{3}{2}\right\rangle$ are not time-reversal symmetric.  For $B_0$ along [001] directions we obtain
\begin{equation}
\frac{1}{T_1}=\frac{(\hbar\omega)^3}{20\rho\pi\hbar^4}\Big(\frac{1}{v_t^5}+\frac{2}{3v_l^5}\Big)2d'^2,
\end{equation}
in agreement with the low-temperature result in ref.~\onlinecite{Ruskov:2013kq}.  

\section{Dephasing from electric field fluctuations}
\label{app:dephasing}

Dephasing of a spin-orbit qubit occurs due to random fluctuations $\hbar\delta\omega(t)$ in the qubit level energy splitting.  In this section we outline expected dephasing rates associated with realistic parameters for the qubit.

\subsection{Charge noise from trap charging/discharging}
First, we estimate the dephasing due to a single fluctuating charge trap (random telegraph signal), assuming that a charge trap can only fluctuate when it is in tunneling proximity to carrier reservoir or gate. The dephasing rate is given by $(T_2^*)^{-1}=(\delta\omega)^2\tau_S/2$, where $\delta\hbar\omega$ is the qubit energy shift when the trap is charged, and $\tau_S$ is the switching time\cite{Bermeister:2014jb}.  Together with its image in the gate, a dipole $e\ell_d$ is created, resulting in a dipole potential $V_{\rm d}=e\ell_d z/4\pi(x^2+z^2)^{3/2}$ when the defect is a distance $\mathbf{R}=\hat{\mathbf{x}}x+\hat{\mathbf{z}}z$ away from the acceptor qubit. The electric field for the dipole potential is $\delta \mathbf{E}=-\nabla V_d=\hat{\mathbf{x}}\delta E_x+\hat{\mathbf{z}}\delta E_z$.

The change $\delta\hbar\omega$ of the Larmor frequency is readily calculated from the extended qubit Hamiltonian (Eq.~\ref{eq:HEsup}), where $\delta E_{x,y}$ couples off-diagonally through the $\alpha E_i+\lambda_{Zi}$ terms, and $\delta E_z$ on-diagonally via the explicit dependence of $\Delta(E_z)$ and $|\lambda_{Zi}(E_z)|$ on $E_z$. Equivalently, the $2\times2$ qubit model can be used, and it is easy to show that 
\begin{equation}
\delta\omega\approx\omega(E_z+\delta E_z)-\omega(E_z)+\frac{2D^2\delta E_{x,y}^2}{\hbar^2\omega(E_z)}.
\end{equation}
For the estimate of dephasing in the main text we assumed $x=50$ nm, and $z=5$ nm giving $\delta E_x\approx 600$ V/m and $\delta E_z\approx 2000$ V/m.  

\subsection{Gate electric field noise}
Gate electric field noise is modeled as a white Johnson voltage noise $v_n(t)$ process applied across the gates producing a randomly fluctuating field of order $E(t)\approx v_n(t)/d_g$ where $d_g\approx 20$ nm is the shortest envisioned distance between gates.  The dephasing rate is given by $(T_2^*)^{-1}=S(\omega=0)$ where $S(\omega)$ is the power spectral density of the random process $\delta\omega(t)$.   Further approximation to $\delta\omega$ from above gives
\begin{equation}
\delta\omega(t)\approx\frac{\partial\omega(E_z)}{\partial E_z}\frac{v_n(t)}{d_g}+\frac{2D^2}{\hbar^2\omega}\Big(\frac{v_n(t)}{d_g}\Big)^2
\end{equation}
We find that the first term is much larger for our qubit and dominates the power spectral density of $\delta \omega(t)$,
\begin{equation}
S(\omega)\approx\Big(\frac{\partial \omega(E_z)}{\partial E_z}\frac{1}{d_g}\Big)^2S_{vv}(\omega).
\end{equation}
The quantity $\partial\omega(E_z)/\partial E_z$ is determined from our analytic model.  Substituting $S_{vv}(\omega)=4k_BTR$ as the white noise power spectrum, $T=1$ K and $R=50$ ohms, we find that the dephasing rate $(T_2^*)^{-1}$ due to intrinsic Johnson voltage noise on the gate is $10^{7}$ times smaller than the dephasing rate from charge noise.

\section{Numerical Kohn Luttinger Calculations}
\label{app:numerics}
All theory preditions in the main text were compared with a numerical solution of the acceptor Hamiltonian including the full spatial dependence of $H(\mathbf{k})$.  The Hamiltonian $H=H_{\rm LK}+H_{E,\rm ion}$, where $H_{\rm LK}=H(\mathbf{k})+H_{\epsilon}+H_{\rm ion}(\mathbf{r})+H_{\rm if}(z)+H_{\rm E}+H_Z$, was computed in the $6\times6$ representation of valence band Bloch states $\left|J,m_J\right\rangle$\cite{Luttinger:1955ee}, $\left|\psi_{m_j}\right\rangle=\sum_{J,m_J}F_{J,m_J}(\mathbf{r})\left|J,m_J\right\rangle$, where $F_{J,m_J}(\mathbf{r})$ are envelope functions.  The first step is to numerically diagonalize $H_{\rm LK}$ using $H(\mathbf{k})=$
\begin{equation}
\left(
\begin{array}{cccccc}
P+Q & -S & R & 0 & -\tfrac{1}{\sqrt{2}}S & \sqrt{2}R\\
-S^\dagger & P-Q & 0 & S & -\sqrt{2}R & \sqrt{\tfrac{3}{2}}S \\
R^\dagger & 0 & P-Q & S & \sqrt{\tfrac{3}{2}}S^\dagger & \sqrt{2}Q \\
0 & R^\dagger & S^\dagger & P+Q & -\sqrt{2}R^\dagger & -\tfrac{1}{\sqrt{2}}S^\dagger \\
-\tfrac{1}{\sqrt{2}}S^\dagger & -\sqrt{2}Q & \sqrt{\tfrac{3}{2}}S & -\sqrt{2}R & P+\Delta_{\rm SO} & 0 \\
\sqrt{2}R & \sqrt{\tfrac{3}{2}}S^\dagger & \sqrt{2}Q & -\tfrac{1}{\sqrt{2}}S & 0 & P+\Delta_{\rm SO}
\end{array}
\right)
\end{equation}
in the $\ket{\tfrac{3}{2},\tfrac{3}{2}}$, $\ket{\tfrac{3}{2},\tfrac{1}{2}}$, $\ket{\tfrac{3}{2},-\tfrac{1}{2}}$, $\ket{\tfrac{3}{2},-\tfrac{3}{2}}$, $\ket{\tfrac{1}{2},\tfrac{1}{2}}$, $\ket{\tfrac{1}{2},-\tfrac{1}{2}}$, basis, with
\begin{align}
P &= \frac{\hbar^2}{2m_0} \gamma_1 (k_x^2+k_y^2+k_z^2), \\
Q &= \frac{\hbar^2}{2m_0} \gamma_2 (k_x^2+k_y^2-2k_z^2), \\
R &= \frac{\hbar^2}{2m_0} \sqrt{3}(-\gamma_2(k_x^2-k_y^2)+2i\gamma_3k_xk_y), \\
S &= \frac{\hbar^2}{2m_0} 2\sqrt{3}\gamma_3(k_x-ik_y)k_z,
\end{align}
$H_{\rm ion}(\mathbf{r})=-e^2/4\pi\epsilon r$, $H_{E}=-eE_zz$, $H_{\rm if}=U_0\Theta(-z)$ with $U_0\rightarrow\infty$ (using a Dirichlet boundary condition), and $H_{Z}$ and $H_{\epsilon}$ from the main text. Here, $k_x=i\tfrac{\partial}{\partial x}$, $k_y=i\tfrac{\partial}{\partial y}$, $k_z=i\tfrac{\partial}{\partial z}$, and $\epsilon_{ij}$ are the strains. The dielectric constant $\epsilon=11.4\epsilon_0$, LK parameters $\gamma_1=4.28$, $\gamma_2=0.375$ and $\gamma_3=1.44$, and $\Delta_{\rm SO}=44$ meV, Bir-Pikus deformation potentials $b'=-1.42$ eV and $d'=-3.7$ eV, and Lande g-factors $g_1=1.07$ and $g_2=0.033$ were obtained from the literature\cite{Neubrand:1978jz,Kopf:1992gj,Lawaetz:1971cm,Bernholc:1977bf}.  The hydrostatic deformation potential $a'$ is not needed since it acts on the valence bands with an identity operator. 

The Hamiltonian $H_{\rm LK}$ was diagonalized by finite differences in Cartesian coordinates for different fields $E_z$ and acceptor depths $d$, to obtain the eigen-energies (eigen-states) $\varepsilon_{\rm LK}$ ($\left|\Psi_{\rm LK}\right\rangle$) of $H_{\rm LK}$.  The inversion asymmetry-induced mixing of states outside the $4\times 4$ subspace is explicitly included due to the explicit representation of electric potentials of the ion, interface, and applied gate field, in the Hamiltonian.  The effect of the $T_d$ symmetry terms is absent in $\left|\Psi_{\rm LK}\right\rangle$ because the local field of the ion is only present in the unit cell containing the ion, which lies outside the validity of the LK approach\cite{Bir:1963iy,Bir:1963cu}. 

The $T_d$ symmetry term was included in the spectrum and eigenstates by diagonalizing $H_{\rm LK}+H_{E,\rm ion}$ in the basis $\left|\Psi_{\rm LK}\right\rangle$.  The $p$ values in $H_{E,\rm ion}$ were taken to reproduce LH-HH couplings known from experiments\cite{Kopf:1992gj} for calculations on bulk acceptors.  We reduced $p$ by a multiplicative prefactor  due to the interface and gate field, which was determined from the reduction of the charge in density at the ion in $\left|\Psi_{\rm LK}\right\rangle$.  

We diagonalized $H_{\rm LK}+H_{E,\rm ion}$ using the first 8 states from $H_{\rm LK}$.  We note that states 5 and 6 contribute very little; even though these predominantly split-off hole states have a non-zero density at the ion (non-zero $T_d$ symmetry correction), they are already 20 meV away (compared to the $\sim 0.2$ meV HH-LH level splitting).  The even higher energy states 7 and higher may be neglected \textit{a priori} at low fields, because they have even higher energies, but more importantly, since their principal quantum number is 2 or higher.  Consequently, the density at the ion, and hence the $T_d$ symmetry coupling, is even smaller. 
\subsection{$H_{\rm if}+H_{E}$: symmetry and parameters for analytic model}
The states $\left|\Psi_{\rm LK}\right\rangle$ experience inversion asymmetry due to explicit presence of the interface and gate field.  We have verified that inversion asymmetry-induced mixing of excited states into the $4\times4$ subspace does not modify the symmetry of $H_{\rm E}$ from the form in the main text, for different acceptor depths, by evaluating $H_{E}=\left\langle \Psi_{m_J',\rm LK} | e\delta \mathbf{E}\cdot\mathbf{r} | \Psi_{m_J,\rm LK}\right\rangle$ using eigenstates $\left|\Psi_{\rm LK}\right\rangle$ calculated non-perturbatively, that is, with applied potentials including the ion, interface, and gate electric field.  This was checked for all static fields $E_z$ in the main text, and for fields $\delta \mathbf{E}$ for all three orthogonal directions.  Values for $\Delta_{\rm if}+\Delta(E_z)$ were obtained from eigen-energies, while values of $\alpha$ were obtained from the off-diagonal elements coupling the LH and HH blocks.  
\subsection{Larmor frequency and EDSR: numerical results}
The LH-HH mixing caused by $p$ in $H_{E,\rm ion}$ changes the Larmor frequency $\hbar\omega$ of our spin qubit $\left|\pm\right\rangle$ dramatically compared with $\left|\Psi_{\pm \rm LK}\right\rangle$.  Numerical calculations of $\hbar\omega$ (Fig.~2C,D, blue squares) are in agreement with the approximate (Fig.~2C,D, black lines) and analytic solutions (Fig.~2C,D, green lines) to the $4\times4$ model. In particular, our numerical calculations show that mixing of states outside the $4\times4$ subspace, and the inclusion of cubic Lande g-factor $g_2$ with values obtained in experiments, does not qualitatively change the behaviour of the qubit splitting.  Therefore, the main electric-field dependence on electric fields is dominated by the LH-HH mixing inside the $4\times4$ subspace.  The numerical results for EDSR (Fig.~2E,F, orange squares) were obtained by transforming our numerical representations of $H_{E}$ and $H_{E,\rm ion}$ into the basis of $H$, using $\left\langle + | e\mathbf{E}\cdot\mathbf{r} | -\right\rangle=U'^\dagger(H_E+H_{E,\rm ion})U'$, where $U'$ is the unitary matrix that diagonalizes $H_{\rm LK}+H_{E,\rm ion}$ in the basis $\left|\Psi_{\rm LK}\right\rangle$.

%

\end{document}